\documentclass[onecolumn,superscriptaddress,aps,pra,showpacs,floatfix]{revtex4}

\usepackage{latexsym}
\usepackage{amsmath}
\usepackage{amssymb}
\usepackage{latexsym}
\usepackage{color}
\usepackage{graphicx}
\usepackage{cancel}
\usepackage{bbm}
\usepackage{bm}
\usepackage{bbm}
\usepackage{maybemath}

\newcommand{\ii}{\mathrm{i}}

\newcommand{\calE}{\mathcal{E}}

\newcommand{\calO}{\mathcal{O}}

\newcommand{\calU}{\mathcal{U}}

\newcommand{\plus}{{\mbox{{\bf{\tiny +}}}}}

\begin{document}

\setlength{\abovecaptionskip}{-5pt} 

\title{Foldy--Wouthuysen Transformation, Scalar Potentials and Gravity}

\author{U. D. Jentschura}

\affiliation{Department of Physics,
Missouri University of Science and Technology,
Rolla, Missouri 65409, USA}

\affiliation{MTA--DE Particle Physics Research Group, 
P.O.Box 51, H-4001 Debrecen, Hungary}

\author{J. H. Noble}

\affiliation{Department of Physics,
Missouri University of Science and Technology,
Rolla, Missouri 65409, USA}

\begin{abstract}
We show that care is required in formulating the nonrelativistic limit of
generalized Dirac Hamiltonians which describe particles and antiparticles
interacting with static electric and/or gravitational fields.  The
Dirac--Coulomb and the Dirac--Schwarzschild Hamiltonians, and the corrections
to the Dirac equation in a non-inertial frame, according to general relativity,
are used as example cases in order to investigate the unitarity of the standard
and ``chiral'' approaches to the Foldy--Wouthuysen transformation, and spurious
parity-breaking terms.  Indeed, we find that parity-violating terms can be
generated by unitary pseudo-scalar transformations (``chiral''
Foldy--Wouthuysen transformations).  Despite their interesting algebraic
properties, we find that ``chiral'' Foldy--Wouthuysen transformations change
fundamental symmetry properties of the Hamiltonian and do not conserve the
physical interpretation of the operators.  Supplementing the discussion, we
calculate the leading terms in the Foldy--Wouthuysen transformation of the
Dirac Hamiltonian with a scalar potential (of the $1/r$-form and 
of the confining radially symmetric linear form), and obtain 
compact expressions for the leading higher-order corrections
to the Dirac Hamiltonian in a non-inertial rotating reference frame (``Mashhoon
term'').
\end{abstract}


\maketitle

% Field theory;
% Relativistic wave equations;
% Quantum electrodynamics;
% Cosmology;
% General relativity and gravitation: 
% Fundamental problems and general formalism;
% Approximation methods; equations of motion.

%
% Introduction
%
\section{Introduction}
\label{sec1}

A (generalized) Dirac Hamiltonian describes the quantum dynamics of a
spin-$1/2$ particle including all relativistic corrections, in the presence of
external fields.  Examples include the the Dirac--Coulomb
Hamiltonian~\cite{FoWu1950,BjDr1964,ItZu1980}, which describes the motion of a
particle bound to the Coulomb field of a nucleus (in the non-recoil
approximation), and the Dirac--Schwarzschild
Hamiltonian~\cite{Ob2001,Je2013,JeNo2013pra}, which describes the motion of a
particle in the curved space-time around a planet or black hole.  The
Foldy--Wouthuysen transformation~\cite{FoWu1950} of the Dirac Hamiltonian
identifies the nonrelativistic limit and the relativistic correction terms.
Let us briefly recall the basic properties.  A static, noninteracting Dirac
particle is described by the Hamiltonian $\beta\,m$, where $\beta$ is the Dirac
$\beta$ matrix with eigenvalues $\pm 1$, and $m$ is the particle mass. The
energy eigenvalues at rest are thus given as $\pm m$, where the positive sign
describes particles, and the negative sign describes antiparticles (we set
$\hbar = c = \epsilon_0 = 1$ in this article).  The nonrelativistic kinetic
energy reads as $\vec p^{\,2}/(2m)$.  Relativistic corrections of order $\vec
p^{\,4}$ and higher in the momenta can be derived from the Foldy--Wouthuysen
transformation. The most important property of the Foldy--Wouthuysen
transformation consists in the disentanglement of the particle and antiparticle
terms and the identification of ``effective'' Hamiltonians which govern the
quantum dynamics of the particles and antiparticles.  

The hierarchy of the terms in the Foldy--Wouthuysen leads to a consistent
perturbative formalism. The so-called Thomas precession of a spinning particle
bound to a Coulomb field~\cite{FoWu1950,BjDr1964,ItZu1980}, or the Fokker
precession of a spinning particle in a gravitational
field~\cite{Fo1920,Kh2001,Kh2008,JeNo2013pra}, can be calculated on the basis of
the Foldy--Wouthuysen transformation. In higher orders,
the calculation of the Foldy--Wouthuysen
transformation can be difficult, notably,
when corrections to the transition current have to be
included~\cite{Je1996,JeNo2013pra}.  Thus, alternative approaches to the
calculation of the Foldy--Wouthuysen transformation have been considered, with
the eventual hope of simplifying the calculation substantially
(see Refs.~\cite{ErKo1958,Ni1998,Ob2001} and references therein).  Typically, the
alternative approaches involve chiral transformations, still unitary, with the
Dirac $\gamma^5$ matrix.  In principle, the alternative approach, which is
based on interesting algebraic identities~\cite{ErKo1958,Ni1998,Ob2001}, has
the potential of fundamentally simplifying the approach to obtaining the
relativistic correction terms and could potentially simplify quantum
electrodynamic bound-state calculations.  Here, we compare the original
Foldy--Wouthuysen approach, and the chiral approach, using a number of example
calculations based on the Dirac Hamiltonians in the Coulomb field,
in the Schwarzschild metric, in a non-inertial frame, and 
for a Dirac particle bound in a scalar potential.
The physical problems studied here are introduced in Sec.~\ref{sec2},
together with a brief explanation of the standard and chiral Foldy--Wouthuysen
transformations, whereas concrete calculations are reserved for
Sec.~\ref{sec3}. Conclusions are drawn in Sec.~\ref{sec5}.
The interpretation of the spin operator
in the context of the chiral method is discussed in~\ref{appa}.

%
% Relativistic Formalism: Outline of the Problem
%
\section{Relativistic Formalism: Outline of the Problem}
\label{sec2}

%
% The Generalized Dirac Hamiltonians Studied
%
\subsection{Generalized Dirac Hamiltonians}
\label{sec2A}

We shall investigate the Dirac equation 
for spin-$1/2$ particles, $\ii \, \partial_t \psi(t, \vec r) = 
H \psi(t, \vec r)$, where $H$ is a (generalized)
Dirac Hamiltonian, $t$ is the time coordinate, and $\psi$ is a four-component
(bispinor) wave function.  The Hamiltonian $H$ is a $(4\times 4)$ matrix in 
spinor space. The free Dirac Hamiltonian
is given by the expression
\begin{equation}
\label{HF}
H_{\rm F} = \vec \alpha \cdot \vec p + \beta m \,.
\end{equation}
The momentum operator is $\vec p$ and the 
mass of the particle is denoted as $m$.
The Dirac matrices $\alpha^i = \gamma^0 \, \gamma^i$ 
(for $i=1,2,3$) and 
$\beta = \gamma^0$ are used in the Dirac representation,
\begin{align}
\label{diracrep}
\gamma^0 =& \; \left( \begin{array}{cc} \mathbbm{1}_{2\times 2} & 0 \\
0 & -\mathbbm{1}_{2\times 2} \\
\end{array} \right) \,,
\quad
\vec\gamma = \left( \begin{array}{cc} 0 & \vec\sigma \\ -\vec\sigma & 0  \\
\end{array} \right) \,,
\quad
\gamma^5 = \left( \begin{array}{cc} 0 & \mathbbm{1}_{2\times 2} \\
\mathbbm{1}_{2\times 2} & 0  \\
\end{array} \right) \,.
\end{align}
These matrices fulfill the relation
$\{ \gamma^\mu, \gamma^\nu \} = 2 \, g^{\mu\nu}$.
in ``West-coast'' conventions with $g^{\mu\nu} = {\rm diag}(1,-1,-1,-1)$.
The Dirac--Coulomb (DC) Hamiltonian is given by the expression
\begin{equation}
\label{HDC}
H_{\rm DC} = \vec \alpha \cdot \vec p + \beta m - \frac{Z\alpha}{r}  \,,
\end{equation}
where $Z$ is the nuclear charge number, $\alpha$ is the 
fine-structure constant, and $r = |\vec r|$ is the distance 
from the nucleus. By contrast, the
scalar potential (SP) multiplies the $\beta$ matrix,
\begin{equation}
\label{HSP}
H_{\rm SP} = \vec \alpha \cdot \vec p + 
\beta \, \left( m - \frac{\lambda}{r} \right) \,,
\end{equation}
where $\lambda$ is a coupling parameter.
The Hamiltonian~\eqref{HSP} can be used as a
rough approximation to nuclear attractive forces,
mediated by meson exchange~\cite{GrMa1996}.
The Dirac Hamiltonian with a radially symmetric, 
linear confining potential is 
\begin{equation}
\label{HLC}
H_{\rm LC} = \vec \alpha \cdot \vec p + 
\beta \, \left( m + \alpha^2 \, r \right) \,,
\end{equation}
where $\alpha$ is a symbolic parameter governing the 
expansion about the nonrelativistic limit:
Namely, for the Foldy--Wouthuysen transformation to 
be physically meaningful, the term $\beta \, m$ in
the Hamiltonian needs to dominant in the nonrelativistic limit.
This implies that the linear confining potential must 
be shallow or the bound particle is always relativistic, 
in which case the Foldy--Wouthuysen transform should not 
be applied.
If we then expand in the momenta $p^i \sim \alpha$ 
and distances $r \sim \alpha^{-1}$, we obtain 
a meaningful expansion about the nonrelativistic limit.

The Dirac--Schwarzschild (DS) Hamiltonian
reads as follows~\cite{Ob2001,Je2013,JeNo2013pra},
\begin{equation}
\label{HDS}
H_{\rm DS} =
\frac12 \, \left\{ \vec\alpha \cdot \vec p,
\left( 1 - \frac{r_s}{r} \right) \right\} +
\beta m \left( 1 - \frac{r_s}{2 r} \right) \,,
\end{equation}
where $\{ ., . \}$ denotes the anticommutator.
The Schwarzschild radius is given as
$r_s = 2 \, G \, M$, where $G$ is Newton's gravitational constant,
and $M$ is the mass of the planet or gravitational centre.
The Dirac Hamiltonian in a non-inertial frame~\cite{HeNi1990,SiTe2005}
reads as 
\begin{equation}
\label{HNF}
H_{\rm NF} = ( 1 + \vec a \cdot \vec r ) \, \beta \, m 
+ \frac12 \, \left\{ 1 + \vec a \cdot \vec r, 
\vec \alpha \cdot \vec p \right\} -
\vec \omega\cdot (\vec L + \tfrac12 \vec \Sigma) \,,
\end{equation}
where $\vec a$ is the acceleration with respect to the 
inertial reference frame. The term involving the 
proper angular rotation frequency $\vec \omega$ is
otherwise known as the Mashhoon term~\cite{Ma1988term}.

%
% Standard Foldy-Wouthuysen Transformation
%
\subsection{Standard Foldy-Wouthuysen Transformation}
\label{sec2B}

For the Foldy--Wouthuysen transformation~\cite{FoWu1950}
one divides a Dirac Hamiltonian $H$
into even and odd (in spinor space) parts,
canonically denoted as $\calE$ and $\calO$.
The even and odd parts of a general operator $H$
are defined as follows,
\begin{equation}
\label{even_odd}
\left\{ H \right\}_{\rm even} 
\equiv \tfrac{1}{2}\, \left(H+\beta H \beta\right) \,,
\qquad
\left\{ H \right\}_{\rm odd} 
\equiv \tfrac{1}{2}\, \left(H-\beta H \beta\right)\,.
\end{equation}
One identifies
\begin{equation}
\calE = \left\{ H \right\}_{\rm even}  \,,
\qquad
\calO = \left\{ H \right\}_{\rm odd} \,.
\end{equation}
For the free Dirac--Hamiltonian~\eqref{HF}, one writes
\begin{equation}
\calE = \beta \, m \,,
\qquad
\calO = \vec\alpha \cdot \vec p \,.
\end{equation}
In the Dirac representation, the $\beta$ matrix anticommutes
with any odd operator, and the term $\beta \, m$, which 
describes a nonrelativistic particle at rest, actually is retained upon
iterating the Foldy--Wouthuysen 
transformation~\cite{FoWu1950,BjDr1964,JeNo2013pra}.
One defines the Hermitian operator $S$ and the unitary transformation
$U$ as follows,
\begin{equation}
\label{USFW}
S = -\ii \frac{\beta}{2 m} \, \calO \,,
\qquad
U = \exp(\ii \, S) \,.
\end{equation}
The Foldy--Wouthuysen transformation is calculated as 
the multi-commutator expansion
\begin{align} 
H' =& \; U \, H \, U^\plus =
\exp(\ii \, S) \, H \, \exp(-\ii S) 
\nonumber\\[2ex]
=& \;  H + \ii \, [S, H] + \frac{\ii^2}{2!} \, [S,[S,H]] + \dots \,,
\end{align}
and it is easy to check that the first commutator 
\begin{equation} 
\ii \, [S, H] \approx 
\ii \, [S, \beta m] = -\calO 
\end{equation} 
generates a term which eliminates the odd operator 
$\calO$ from the transformed Hamiltonian $H'$.
However, many more (possibly odd) terms are generated
by the higher-order terms in the Foldy--Wouthuysen 
Hamiltonian, which may have to be eliminated using 
subsequent transformations $U$, $U'$, $U''$, and so on,
until all odd operators are eliminated up to 
a given order in the perturbative expansion. 
One usually defines a perturbative parameter
(e.g., the power of the momentum operator, or, 
in classical terms, the velocity of the particle
expressed in units of the speed of light)
and keeps terms only up to a specified order in 
this parameter.

%
% Chiral Foldy-Wouthuysen Transformation
%
\subsection{Chiral Foldy-Wouthuysen Transformation}
\label{sec2C}

In Refs.~\cite{ErKo1958,Ni1998,Ob2001},
an alternative variant of the ``Foldy-Wouthuysen'' transformation is proposed,
which we would like to refer to as the ``chiral'' Foldy--Wouthuysen 
transformation because it has a number of interesting 
algebraic properties, and because it actually contains 
``chiral'' $\gamma^5$ matrices.
At face value, the proposed method leads to an exact 
separation of the input Hamiltonian into even and odd contributions,
which are straightforward to expand in the perturbative parameters.
The proposed rotation contains a combination of two unitary transformations,
which are both chiral,
\begin{equation}
\label{U1U2}
U=U_2 \, U_1 \,,
\quad
U_1=\frac{1}{\sqrt{2}}\left(1+J\,\Lambda\right)\,,
\quad
U_2=\frac{1}{\sqrt{2}}\left(1+\beta J\right)\,.
\end{equation}
The operators $\Lambda$ and $J$ are Hermitian roots of unity,
\begin{equation}
\label{important}
\Lambda =; \frac{H}{\sqrt{H^2}}\,,
\quad
\Lambda^\plus = \Lambda \,,
\quad
J = \ii\,\gamma^5\,\beta\,,
\quad
J^\plus = J \,,
\quad
\Lambda^2 = J^2 = 1 \,.
\end{equation}
Of course, $H$ is the Hamiltonian that we are trying to transform,
and it is understood that the square root of its square,
$\sqrt{H^2}$, can be expanded easily 
in terms of the perturbative parameters.
For the chiral transformation to work, it is essential that
\begin{equation}
\label{important1}
\left\{\Lambda, J\right\} = \left\{H,J\right\} =
\left\{\sqrt{H^2}, J\right\} = 0 \,.
\end{equation}
The following proof of the unitarity of $U$
depends on the property~\eqref{important1},
\begin{align}
\label{eq1.6}
U\,U^+ = & \; U_2\,U_1\,U_1^+\,U_2^+
= U_2 \, \tfrac{1}{2}(1+J\Lambda)(1+\Lambda J) \, U_2
\nonumber\\
=& \; \tfrac{1}{2} \, U_2(2+J\Lambda+\Lambda J)U_2^+
\nonumber\\
=& \; \tfrac{1}{4} \, (1+\beta J)(2+J\Lambda+\Lambda J)(1+J\beta)
\nonumber\\
=& \; \tfrac{1}{4} \, (2+2\beta J+J\Lambda+\beta\Lambda+\Lambda J
\nonumber\\
&
+\beta J\Lambda J
+2J\beta +2\beta JJ\beta+J\Lambda J\beta
\nonumber\\
&\;
+\beta\Lambda J\beta
+\Lambda JJ\beta+\beta J\Lambda JJ\beta) \,.
\end{align}
Taking notice of the properties~\eqref{important}
and~\eqref{important1}, as well as the relation
$J\beta= -\beta J$, one can show that
all terms in Eq.~\eqref{eq1.6} mutually cancel except for
\begin{equation}
U \, U^\plus = \tfrac{1}{4}\,(2+2\beta J \, J \, \beta) =1\,.
\end{equation}
The following surprisingly simple and elegant result~\cite{Ob2001,Ni1998}
(after some manipulations which we give in detail)
is central to the chiral Foldy--Wouthuysen transformation,
\begin{align}
\label{eq1.13}
U\,H\,U^+ = & \; U_2U_1\,H\,U_1^+U_2^+
\nonumber\\
=&\; \tfrac{1}{2} \, U_2(H+J\Lambda H+H\Lambda J+J\Lambda H\Lambda J)U_2^+
\nonumber\\
=& \; U_2\,J\Lambda H\,U_2^+
=\tfrac{1}{2} \, (1+\beta J)J\Lambda H(1+J\beta)
\nonumber\\
=& \; \tfrac{1}{2} \, (J\Lambda H+\beta\Lambda H
+\Lambda H\beta-\beta\Lambda H\beta J)
\nonumber\\
=& \; \tfrac{1}{2} \, \beta (\sqrt{H^2}+ \beta \, \sqrt{H^2} \, \beta)
+ \tfrac{1}{2}(\sqrt{H^2}-\beta\sqrt{H^2}\beta)J
\nonumber\\
=& \; \left\{ \sqrt{H^2} \right\}_{\rm even} \; \beta+
\left\{ \sqrt{H^2} \right\}_{\rm odd} \; J\,.
\end{align}
The operator $J$ is odd in 
spinor space, and both terms in the last line of Eq.~\eqref{eq1.13}
constitute even expressions in spinor space.
One might thus assume that the chiral Foldy--Wouthuysen
transformation considerably simplifies
the identification of the nonrelativistic 
limit of generalized Dirac Hamiltonians:
The transformed Hamiltonian is even in spinor space,
and the (relativistic) expansion of the 
square root of the square of the ``input'' Hamiltonian $H$
is generally accomplished easily.
The chiral Foldy--Wouthuysen transformation might thus 
completely eliminate the need for the 
complicated evaluation of multiple commutators,
as would otherwise be the case for the standard
Foldy--Wouthuysen transformation.
Conceivably, the chiral Foldy--Wouthuysen transformation
would thus lead to a much simplified identification
of higher-order terms in the Breit--Pauli Hamiltonian
for atoms interacting with external fields,
and quantization radiation fields,
where considerable effort has been invested 
in recent years~\cite{Pa2004,Pa2005,JeCzPa2005,PaCzJeYe2005,Pa2007} 
in the identification of the 
general $(Z\alpha)^6$ higher-order correction terms
for bound systems.

%
% Application to the Hamiltonians
%
\section{Application to the Hamiltonians}
\label{sec3}

%
% Free Dirac and Dirac--Coulomb Hamiltonian
%
\subsection{Free Dirac and Dirac--Coulomb Hamiltonian}
\label{sec3A}

For the free Dirac Hamiltonian 
$H_{\rm F} = \vec \alpha \cdot \vec p + \beta m$
defined in Eq.~\eqref{HF}, it is easy to verify that 
\begin{equation}
\label{free_yes}
\{ \vec \alpha \cdot \vec p, \; \ii \, \gamma^5 \, \beta \} = 0 \,,
\quad
\{ \beta m, \; \ii \,\gamma^5 \, \beta \} = 0 \,,
\quad
\{ H_F, J \} = 0 \,.
\end{equation}
The conditions for the application of the chiral 
Foldy--Wouthuysen transformation are thus fulfilled.
One calculates
\begin{equation}
H_{\rm F}^2 = \vec p^{\, 2} + m^2 \,,
\qquad
\sqrt{ H_{\rm F}^2 } = m + \frac{\vec p^{\, 2}}{2 \, m} - 
\frac{\vec p^{\, 4}}{8 \, m^3} + \dots \,,
\end{equation}
where the expansion is carried out in ascending powers of 
the momenta, and thus 
$\left\{ \sqrt{ H_{\rm F}^2 } \right\}_{\rm even}
=\sqrt{ H_{\rm F}^2 }$ while
$\left\{ \sqrt{ H_{\rm F}^2 } \right\}_{\rm odd}$ vanishes.
Formula~\eqref{eq1.13} immediately leads to the result
\begin{equation} 
H^{\rm (CFW)}_{\rm F} =
\beta \, \left(  m + \frac{\vec p^{\, 2}}{2 \, m} - 
\frac{\vec p^{\, 4}}{8 \, m^3} + \dots \right) \,,
\end{equation} 
where by the superscript CFW we denote the result
of the chiral Foldy--Wouthuysen transformation.
It is well known~\cite{FoWu1950,BjDr1964} that the 
standard Foldy--Wouthuysen (SFW) transformation leads to the 
same result,
\begin{equation} 
H^{\rm (SFW)}_{\rm F} = H^{\rm (CFW)}_{\rm F} =
\beta \, \sqrt{ \vec p^{\, 2} + m^2 } \,,
\end{equation}
where we indicate the nonperturbative nature
(in powers of the momenta) in the transformed result
(see Chap.~4 of Ref.~\cite{BjDr1964}).
The Dirac--Coulomb Hamiltonian $H_{\rm DC}$
contains that Coulomb potential $V = -Z\alpha/r$,
\begin{equation}
H_{\rm DC} = \vec \alpha \cdot \vec p + \beta m - \frac{Z\alpha}{r}  \,.
\end{equation}
The first two terms fulfill the condition~\eqref{important1}
[see Eq.~\eqref{free_yes}], but the third term fulfills
\begin{equation}
\label{comm_term}
[ V, J ] =
\left[ -\frac{Z\alpha}{r}, \ii \, \gamma^5 \, \beta \right] = 0 \,.
\end{equation}
instead of the corresponding relation with the anticommutator.
Therefore, strictly speaking, the conditions for the 
application of the chiral Foldy--Wouthuysen transformation
are not fulfilled. However, there are several known example
cases in physics and mathematics where the application of a mathematical method leads to 
consistent results, even if the corresponding conditions,
strictly speaking,
are not fulfilled. A rather famous, yet distant, example is the 
use of asymptotic expansions in the non-asymptotic regime, 
which lead to perfectly consistent results if they are
combined with suitable resummation 
prescriptions~\cite{We1989,JeMoSo1999,JeMoSo2001pra}.

It is thus more than an academic exercise to apply
the formalism of the chiral Foldy--Wouthuysen transformation
to the Dirac--Coulomb Hamiltonian, and to 
investigate the results.
We first square the Dirac--Coulomb Hamiltonian to find
\begin{equation}
H_{\rm DC}^2= m^2 + \vec p^{\,2} -
\left\{ \vec\alpha\cdot \vec p, \frac{Z\alpha}{r}\right\}
-2\beta m\frac{Z\alpha}{r}
+\frac{Z^2\alpha^2}{r^2} \,.
\end{equation}
It is easy to expand the square root of $H_{\rm DC}^2$ to second
order in $Z\alpha$,
\begin{equation}
\sqrt{H_{\rm DC}^2}\approx 
m+\frac{\vec p^{\,2}}{2m}
-\frac{\ii \, Z\alpha}{2m \, r^3}  \, \vec\alpha \cdot \vec r
-\frac{Z\alpha}{mr}\vec\alpha\cdot\vec p
-\beta\frac{Z\alpha}{r} 
\end{equation}
We here suppress the term $\frac{Z^2\alpha^2}{2mr^2}$
on the right-hand side because it is of order 
$(Z\alpha)^4 \, m$ (we recall that for atomic systems,
$\vec p \sim Z\alpha m$ and $r \sim 1/(Z\alpha m)$,
see Refs.~\cite{Pa1993,JePa1996}).
The two terms in Eq.~\eqref{eq1.13} are then identified as follows,
\begin{subequations}
\begin{align}
\left\{ \sqrt{H_{\rm DC}^2} \right\}_{\rm even} \,\beta=& \;
\beta \left( m+\frac{p^2}{2m}\right)-\frac{Z\alpha}{r} \,,
\\
\left\{ \sqrt{H_{\rm DC}^2} \right\}_{\rm odd} \, J=&\;
\frac{Z\alpha}{2m} \, \frac{\beta \, \vec\Sigma\cdot\vec r}{r^3}
-\ii \, \frac{Z\alpha}{mr} \, \beta \, \vec\Sigma\cdot\vec p \,.
\end{align}
\end{subequations}
Here, the vector of $(4 \times 4)$-spin matrices 
has the representation $\Sigma^i = \gamma^5 \, \gamma^0 \, \gamma^i$.
The chiral Foldy--Wouthuysen transformation of the 
Dirac--Coulomb Hamiltonian,
\begin{equation}
\label{HCFW_DC}
H^{\rm (CFW)}_{\rm DC} =
\beta \left( m+ \frac{\vec p^{\,2}}{2m} \right) - \frac{Z\alpha}{r}
+\frac{Z\alpha}{2m} \, \frac{\beta\,\vec\Sigma\cdot\vec r}{r^3}
-\ii\frac{Z\alpha}{mr} \beta \, \vec\Sigma\cdot\vec p \,,
\end{equation}
is different from the result of the 
standard Foldy--Wouthuysen transformation 
(see Refs.~\cite{FoWu1950,BjDr1964,ItZu1980,JePa1996,Je1996}),
which reads as 
$H^{\rm (SFW)}_{\rm DC} \approx
\beta \left( m+ \frac{\vec p^{\,2}}{2m} \right) - \frac{Z\alpha}{r}$
to second order in the $(Z\alpha)$-expansion, and
\begin{equation}
\label{HSFW_DC}
H^{\rm (SFW)}_{\rm DC}=
\beta \left( m+ \frac{\vec p^{\,2}}{2m} - 
\frac{\vec p^{\,4}}{8 m^3} \right) - \frac{Z\alpha}{r} 
+ \frac{\pi \, Z\alpha}{2 m^2} \, \delta^{(3)}(\vec r) 
+ \frac{Z\alpha}{4 m^2 r^3} \, \vec \Sigma \cdot \vec L \,,
\end{equation}
to fourth order in $Z\alpha$. In the latter result,
the zitterbewegung term, and the Thomas precession
term (spin-orbit coupling) are consistently taken into account.
One may observe that the chiral Foldy--Wouthuysen transformation
fails to reproduce the known result for the Dirac--Coulomb Hamiltonian,
as a consequence of the fact that the condition for its application
is not fulfilled, as shown in Eq.~\eqref{comm_term}.
Because the chiral Foldy--Wouthuysen is a one-step, ``exact''
process, there is no possibility that the spurious
terms in Eq.~\eqref{HCFW_DC} are eliminated upon 
the consideration of ``higher orders'' in the transformation.
One may also point out that
the last term in the transformed Hamiltonian~\eqref{HCFW_DC} 
is not even Hermitian, as a consequence of an application
of the chiral transformation beyond its range of applicability.
The manifest failure of the chiral method for 
the phenomenologically important case of the 
Dirac--Coulomb Hamiltonian indicates that the range of applicability 
of chiral transformation might be somewhat limited.

%
% The Model Problem
%
\subsection{Dirac Hamiltonian with Scalar ($1/r$)--Potential}
\label{sec3B}

In contrast to the Dirac--Coulomb Hamiltonian,
the Dirac Hamiltonian with a scalar potential,
given in Eq.~\eqref{HSP},
\begin{equation}
H_{\rm SP} = \vec \alpha \cdot \vec p + 
\beta \, \left( m - \frac{\lambda}{r} \right) \,,
\end{equation}
fulfills the criteria for the application
of the chiral transformation, because 
the anticommutator $\{ H_{\rm SP}, J \}$ 
vanishes. We recall that the name ``scalar potential''
is derived from the covariant representation 
of the corresponding Dirac equation,
$(\ii \gamma^\mu \, \partial_\mu - m + \lambda/r) \, \psi(t, \vec r) = 0$,
where the potential enters as a Lorentz scalar
(the Einstein summation convention is used 
for the sum over $\mu$).
We obtain for the square of the Hamiltonian,
\begin{equation}
H^2_{\rm SP} \approx \vec p^{\,2} + m^2 
+ \beta \, \left[  \vec \alpha \cdot \vec p, \, \frac{\lambda}{r} \right] 
- 2 m \frac{\lambda}{r}  \,,
\end{equation}
where we ignore higher-order terms of order $\lambda$
and terms beyond second order in the momentum.
To second order in the momenta and first order in $\lambda$, we thus have
\begin{equation}
\sqrt{ H^2_{\rm SP} } \approx m + \frac{\vec p^{\,2}}{2 m} 
+ \frac{\ii \, \lambda \, \beta}{2 m} \, \frac{ \vec \alpha \cdot \vec r}{r^3}
- \frac{\lambda}{r} \,.
\end{equation}
For the scalar potential,
the two terms in Eq.~\eqref{eq1.13} are thus identified as follows,
\begin{subequations}
\begin{align}
\left\{ \sqrt{H_{\rm SP}^2} \right\}_{\rm even} \,\beta=& \;
\beta \left( m+\frac{p^2}{2m} - \frac{\lambda}{r} \right) \,,
\\
\left\{ \sqrt{H_{\rm SP}^2} \right\}_{\rm odd} \, J=&\;
-\frac{\lambda}{2 m} \, \frac{ \vec \Sigma \cdot \vec r}{r^3} \,.
\end{align}
\end{subequations}
The result of the chiral transformation of the 
Dirac Hamiltonian with a scalar potential thus is
\begin{equation}
\label{HCFW_SP}
H^{\rm (CFW)}_{\rm SP}=
\beta \left( m+\frac{\vec p^{\,2}}{2m} - \frac{\lambda}{r} \right) 
- \frac{\lambda}{2 m} \, \frac{ \vec \Sigma \cdot \vec r}{r^3} \,.
\end{equation}
The occurrence of the first term proportional to the $\beta$ 
matrix had to be expected, but the second (pseudo-scalar) term 
breaks parity, even though the ``input'' Hamiltonian
$H_{\rm SP}$ is parity invariant.
The Hamiltonian~\eqref{HCFW_SP} is Hermitian,
as it should be, because the chiral transformation is unitary 
in the case of a scalar potential 
(we have $\{ H_{\rm SP}, J \} = 0$).

Within the standard Foldy--Wouthuysen 
procedure outlined in Sec.~\ref{sec2B},  the 
transformation of the Dirac Hamiltonian with a scalar 
potential follows the lines of the standard 
method outlined in Sec.~\ref{sec2B}.
To second order in the momenta, the result reads as
$H^{\rm (SFW)}_{\rm SP} \approx
\beta \left( m+\frac{\vec p^{\,2}}{2m} - \frac{\lambda}{r} \right)$
and needs to be compared with Eq.~\eqref{HCFW_SP}.
To fourth order in the momenta, we have
\begin{align}
\label{HSFW_SP}
H^{\rm (SFW)}_{\rm SP}=& \;
\beta \left( m+\frac{\vec p^{\,2}}{2m} - 
\frac{\vec p^{\,4}}{8 m^3} - \frac{\lambda}{r} 
+\frac{\lambda}{4m^2}\left\{\vec p^{\,2},\frac{1}{r}\right\}
-\frac{\pi \, \lambda}{2 m^2} \; \delta^{(3)}(\vec r) 
- \frac{\lambda \vec \Sigma \cdot \vec L }{4 m^2 r^3} \right) \,.
\end{align}
We obtain the spin-orbit coupling term
(proportional to $\vec \Sigma \cdot \vec L$) and the
zitterbewegung term (proportional to the Dirac-$\delta$).
All the terms in the result~\eqref{HSFW_SP} 
are parity-invariant, as they should be, 
because we started from the parity-invariant 
Hamiltonian~\eqref{HSP}. 
(Both the spin as well as the orbital angular momenta
are pseudo-vectors).
The scalar character of the potential
is manifest in the universal prefactor 
$\beta$ in the Foldy--Wouthuysen transformed 
Hamiltonian~\eqref{HSFW_SP}, which ensures
that both particles as well as antiparticles 
are attracted by the potential proportional to
$\beta \lambda/r$. This is in contrast to the Coulomb potential,
which is attractive for electrons (positive-energy 
eigenstates of the Dirac--Coulomb  Hamiltonian,
but repulsive for negative-energy solutions
to the Dirac--Coulomb  Hamiltonian).
The physical interpretation of the $\vec \Sigma$ spin 
matrices is preserved under the standard approach~\cite{FoWu1950}.

% 
% Dirac Hamiltonian with Scalar Confining Potential
% 
\subsection{Dirac Hamiltonian with Scalar Confining Potential}
\label{sec3C}

We consider the Hamiltonian~\eqref{HLC}
\begin{equation}
H_{\rm LC} = \vec \alpha \cdot \vec p + \beta (m + \alpha^2\, m^2 \, r) \,,
\end{equation}
where we distinguish the vector $\vec\alpha = \gamma^0 \, \vec\gamma$ 
of Dirac matrices from the coupling parameter $\alpha$.
The potential $W = \beta \alpha^2\, m^2 \, r$ 
anticommutes with $J$,
\begin{equation}
\label{comm_term2}
\left\{ W, \, J \right\} =
\left\{ \beta \, \alpha^2 \, m^2 \, r , \ii \, \beta \, \gamma^5 \right\} = 0 \,,
\end{equation}
and the condition for the applicability of the chiral method 
is thus fulfilled. One finds
\begin{equation}
H_{\rm LC}^2= m^2 + \vec p^{\,2} - \beta \,
\left[ \vec\alpha\cdot \vec p, \alpha^2 \, m^2 \, r\right]
+ 2 \alpha^2 \, m^3 \, r + \alpha^4 m^4 \, r^2 \,.
\end{equation}
In the regime where $p^i \sim \alpha$, and $r \sim 1/\alpha$,
one finds to second order in $\alpha$,
\begin{align}
\sqrt{H_{\rm LC}^2}\approx & \; m+\frac{\vec p^{\,2}}{2m}
+ \frac{\beta m}{2} \left[ \vec\alpha\cdot \vec p, \alpha^2 \, r \right]
+ \alpha^2 \, m^2 \, r + \tfrac12 \, \alpha^4 m^3 \, r^2 
\\[2ex]
=& \;
m+\frac{\vec p^{\,2}}{2m}
-\frac{\ii \, \alpha^2}{2 \, r}  \, 
\beta \, m \, \vec\alpha \cdot \vec r
+ \alpha^2 \, m^2 \, r + \tfrac12 \, \alpha^4 m^3 \, r^2 \,,
\end{align}
and thus 
\begin{subequations}
\begin{align}
\left\{ \sqrt{H_{\rm LC}^2} \right\}_{\rm even} \,\beta=& \;
\beta \left( m+\frac{p^2}{2m} + \alpha^2 \, m^2 \, r + 
\tfrac12 \, \alpha^4 m^3 \, r^2 \right) \,,
\\
\left\{ \sqrt{H_{\rm DC}^2} \right\}_{\rm odd} \, J=&\;
-\frac{\alpha^2 \,m}{2} \, \frac{\vec\Sigma\cdot\vec r}{r} \,.
\end{align}
\end{subequations}
The chiral Foldy--Wouthuysen transformation of the 
Dirac--Coulomb Hamiltonian with a scalar, confining potential
thus reads as 
\begin{equation}
\label{HCFW_LC}
H^{\rm (CFW)}_{\rm LC}=
\beta \left( m+ \frac{\vec p^{\,2}}{2m} +
\alpha^2 \, m^2 \, r + \tfrac12 \, \alpha^4 m^3 \, r^2 \right)
- \frac{\alpha^2 \, m}{2} \, \frac{\vec\Sigma\cdot\vec r}{r} \,.
\end{equation}
Two last terms in Eq.~\eqref{HCFW_LC}
are again pseudo-scalar and break parity 
(spin is a pseudo-vector, while $\vec r$ is a vector).
Within the standard approach to the Foldy--Wouthuysen
transformation, the 
spin-orbit coupling here enters at order $\alpha^3$,
and we have
\begin{equation}
\label{HSFW_LC}
H^{\rm (SFW)}_{\rm LC}=
\beta \left( m + \frac{\vec p^{\,2}}{2m} + \alpha^2 \, m^2 \, r 
- \frac{\alpha^2}{4} \, \left\{ r, \vec p^{\,2} \right\}
- \frac{\alpha^2}{4 \, r} 
- \frac{\alpha^2 \, m}{4} \, \frac{\vec\Sigma \cdot \vec L}{r} \right) \,.
\end{equation}
Again, we obtain an anticommutator term of the 
binding potential with the operator $\vec p^{\,2}$, 
and we recover full particle-antiparticle symmetry
(overall prefactor $\beta$).

%
% Dirac-Schwarzschild Hamiltonian
%
\subsection{Dirac-Schwarzschild Hamiltonian}
\label{sec3D}

We now turn our attention to the 
Dirac--Schwarzschild Hamiltonian~\eqref{HDS},
which we recall for convenience,
\begin{equation}
\label{HDS2}
H_{\rm DS} =
\frac12 \, \left\{ \vec\alpha \cdot \vec p,
\left( 1 - \frac{G M}{2 r} \right) \right\} +
\beta m \left( 1 - \frac{G M}{r} \right) \,,
\end{equation}
replacing the Schwarzschild radius from 
Eq.~\eqref{HDS} according to $r_s = 2 \, G \, M$.
With the identification $\lambda = G \, M$, 
the scalar ``potential'' in the mass term in Eq.~\eqref{HDS2}
is exactly equal to the scalar potential in 
Eq.~\eqref{HSP}, but the kinetic term also is affected
in Eq.~\eqref{HDS2}.
Because $\{ H_{\rm DS}, J \} = 0$, the 
conditions for the application of the 
chiral Foldy--Wouthuysen method are fulfilled.

The result from Eq.~(31) of Ref.~\cite{Ob2001},
rewritten in terms of the gravitational constant $G$ and 
the mass $M$ of the planet, reads as
\begin{equation}
\label{HCFW_DS}
H_{\rm DS}^{\rm (CFW)} = \beta\left(m+\frac{\vec p^{\,2}}{2m} 
- \frac{G m M}{r}
+ \frac{2 \pi G M}{m} \, \delta^{(3)}(\vec r)
+ \frac{G M}{m} \frac{\vec\Sigma\cdot \vec L}{r^3} \right)
- \frac{G M }{2 m}\frac{\vec\Sigma\cdot\vec r}{r^3} \,.
\end{equation}
As compared to Eq.~(31) of Ref.~\cite{Ob2001},
we here leave the leading gravitational interaction
term proportional to $G m M/r$ in unexpanded form
(it is written as $m \, \vec g \cdot \vec x$ for a small
displacement $\vec x$ from the position $\vec r$, 
where $\vec g$ is the acceleration due to gravity).
We thus confirm that the formalism of the 
chiral Foldy--Wouthuysen transformation has been
consistently applied in Ref.~\cite{Ob2001} in order to 
obtain the result given in Eq.~\eqref{HCFW_DS}.
The term proportional to                               
$\vec \Sigma \cdot \vec r$ in Eq.~\eqref{HCFW_DS}
indicates that the                 
symmetry of the problem has been altered: The ``input''
Hamiltonian~\eqref{HDS} is parity-even, while the
term proportional to $\vec \Sigma \cdot \vec r$
in the result of the chiral Foldy--Wouthuysen transformation~\eqref{HCFW_DS}
constitutes a pseudo-scalar.
The parity-breaking term is spurious and indicates that the
physical interpretation of the spin operator
$\vec \Sigma$ has been altered~\cite{SiTe2005}.

By contrast, the standard Foldy--Wouthuysen transformation 
leads to a different result~\cite{DoHo1986,SiTe2005,JeNo2013pra}, 
given recently in manifestly Hermitian form~\cite{JeNo2013pra},
\begin{equation}
\label{HSFW_DS}
H^{\rm (SFW)}_{\rm DS} \!=\!
\beta \!\left( m \!+\! \frac{\vec p^{\,2}}{2 m} 
\!-\!
\frac{\vec p^{\,4}}{8 m^3}
\!-\! \frac{G m M}{r}
\!-\! \frac{3 G M}{4 m} \left\{ \vec p^{\,2}, \frac{1}{r} \right\}
\!+\! \frac{3 \pi G M \delta^{(3)}(\vec r)}{2 m} 
\!+\! \frac{3 G M \, \vec\Sigma \cdot \vec L}{4 m r^3} \right) \!.
\end{equation}
The leading gravitational interaction is consistently 
obtained with the prefactor $\beta$ in both 
approaches~\eqref{HCFW_DS} and~\eqref{HSFW_DS}.
The gravitational spin-orbit coupling term in Eq.~\eqref{HSFW_DS} 
is in agreement
with classical physics~\cite{Fo1920,Kh2001,Kh2008}.

%
% Dirac Hamiltonian in a Non-Inertial Frame
%
\subsection{Dirac Hamiltonian in a Non--Inertial Frame}
\label{sec3E}

We recall the Dirac Hamiltonian in a non-inertial 
frame from Eq.~\eqref{HNF},
\begin{equation}
H_{\rm NF} = ( 1 + \vec a \cdot \vec r ) \, \beta \, m 
+ \frac12 \, \left\{ 1 + \vec a \cdot \vec r, \,
\vec \alpha \cdot \vec p \right\} -
\vec \omega\cdot (\vec L + \tfrac12 \vec \Sigma)\,.
\end{equation}
This Hamiltonian is valid for an accelerated frame 
of reference accelerated with a uniform acceleration
vector $\vec a$. Because of the spatially uniform 
acceleration, the magnitude of the coordinate 
$\vec r$ is not bound by any dimension of the system,
It is therefore indicated to carry out the perturbative
program of the Foldy--Wouthuysen transformation
as an expansion in powers of the parameter $\xi$,
where $\vec p \sim \xi$ and $\vec r \sim 1$, i.e.,
the spatial coordinate is treated as a quantity of order
unity. Furthermore, in all calculations below,
we keep the acceleration vector $\vec a$ and 
the angular rotation frequency vector $\vec\omega$ 
only to first order.

Using the operator identity
\begin{equation}
\{ A, B \}^2 - 2 \{ A^2, B^2 \} = 3 \, [A, B] \, [B, A] \,,
\end{equation}
which is valid provided $[A, [A,B]] = [B, [A,B]] = 0$,
with $A = \frac12 \, (1 + \vec a \cdot \vec r)$ 
and $B = \vec \alpha \cdot \vec p$,
one verifies the relation
\begin{equation}
H_{\rm NF}^2 \approx
( 1 + 2 \, \vec a \cdot \vec r ) \, m^2 
+ \frac12 \, \left\{ 1 + 2 \, \vec a \cdot \vec r, \, \vec p^{\,2} \right\}
+ \ii \beta m\, \vec \alpha \cdot \vec a 
- 2 \beta m \vec \omega\cdot (\vec L + \tfrac12 \vec \Sigma) \,,
\end{equation}
where quadratic terms in the parameter $\vec a$ and 
$\omega$ have been neglected. Therefore,
\begin{equation}
\sqrt{H_{\rm NF}^2} \approx
m \, \left( 1 + \vec a \cdot \vec r \right)
+ \frac{\vec p^{\,2}}{2 m} 
+ \frac{1}{2m} \, \left\{ \vec a \cdot \vec r, \vec p^{\,2} \right\}
+ \ii \frac{\beta}{2 m} \, \vec \alpha \cdot \vec a 
- \beta \vec \omega\cdot (\vec L + \tfrac12 \vec \Sigma) \,.
\end{equation}
The chiral Foldy--Wouthuysen transform therefore reads as
follows,
\begin{equation}
\label{HCFW_NF}
H^{\rm (CFW)}_{\rm NF} = 
\beta \, \left( m \, \left( 1 + \vec a \cdot \vec r \right)
+ \frac{\vec p^{\,2}}{2 m} 
+ \frac{1}{2m} \, \left\{ \vec a \cdot \vec r, \vec p^{\,2} \right\} \right)
- \vec \omega\cdot (\vec L + \tfrac12 \vec \Sigma)
+ \frac{1}{2 m} \, \vec \Sigma \cdot \vec a \,.
\end{equation}
The pseudo-scalar term proportional to 
$ \vec \Sigma \cdot \vec a$ has the same structure
as the corresponding term in Eq.~\eqref{HCFW_DS},
if we identify the acceleration due to gravity 
$\vec g = -G M \vec r/r^3$ with $\vec a$.

The standard Foldy--Wouthuysen transformation 
leads to a different result. We would like to 
investigate terms up to fourth order in the momenta.
In the sense of Sec.~\ref{sec2B},
one then has to perform a {\em four-fold} Foldy--Wouthuysen
transformation [four iterated transformations of the 
form given in Eq.~\eqref{USFW}], to obtain the result
\begin{align}
\label{HSFW_NF}
H^{\rm (SFW)}_{\rm NF} =& \; \beta \left( m + \frac{\vec p^{\,2}}{2 m} - 
\frac{\vec p^{\,4}}{8 m^3} 
+ \vec a \cdot \vec r
+ \frac{1}{4 m} \, \{ \vec p^{\,2}, \vec a \cdot \vec r \}
- \frac{1}{16 m^3} \, \{ \vec p^{\,4}, \vec a \cdot \vec r \} \right)
\nonumber\\[1.33ex]
& \; + \beta \, \left[ \vec\Sigma \cdot ( \vec a \times \vec p ) \right]
\, \left( \frac{1}{4 m} - \frac{\vec p^{\,2}}{16 m^3} \right)
- \vec \omega \cdot \left( \vec L + \tfrac12 \, \vec \Sigma \right) \,.
\end{align}
This result fully conserves parity.
It is in agreement with Eq.~(20) of Ref.~\cite{HeNi1990}
but adds higher-order terms (in the momenta).
In order to verify consistency with 
Eq.~(20) of Ref.~\cite{HeNi1990},
one notices that $[p^i, [p^j, \vec a \cdot \vec r]] = 0$,
where the $p^i$ denote the Cartesian components of the 
momentum vector.
Our result~\eqref{HSFW_NF} also is in agreement with 
various other recent investigations~\cite{SiTe2005,SiTe2007,ObSiTe2009}
in appropriate limits, and with Eq.~(5.14) of Ref.~\cite{ObSiTe2011}. 
The results in Refs.~\cite{SiTe2005,SiTe2007,ObSiTe2009,ObSiTe2011}
are obtained within a conceptually different 
approach to the Foldy--Wouthuysen transformation.
The result~\eqref{HSFW_NF} confirms that the 
Mashhoon term receives no relativistic 
corrections up to the relative order $\vec p^4$,
and indicates the leading fourth-order (in the momenta) relativistic corrections
to the spin-orbit coupling in the accelerated frame,
in compact form.

%
% Conclusions
%
\section{Conclusions}
\label{sec5}

We have contrasted the standard and the chiral method for the Foldy--Wouthuysen
transformation of generalized Dirac Hamiltonians.  The chiral method is based
on very interesting, surprising and perhaps, quite fascinating  operator
identities discussed in Sec.~\ref{sec2C}.  A somewhat ``hidden'' assumption of
the chiral method implies that the input Hamiltonian anticommutes with the
chiral operator $J = \ii \gamma^5 \beta$.  We confirm that the chiral method
leads to a consistent result for the free Dirac Hamiltonian (see
Sec.~\ref{sec3A}).  However, the Dirac-Coulomb Hamiltonian (see
Sec.~\ref{sec3A}), for scalar $(1/r)$-potentials and 
scalar confining potential (see Secs.~\ref{sec3B} and~\ref{sec3C}), as well as for the
Dirac--Schwarzschild Hamiltonian (see Sec.~\ref{sec3D}), and for a Dirac
particle moving in an accelerated reference frame (see Sec.~\ref{sec3E}). 

The eventual goal of the ``chiral'' transformation is the calculation of an
``exact'' Foldy--Wouthuysen transformation, which disentangles the particle and
antiparticle degrees of freedom, without the restrictions set forth by a
perturbative formalism. Such calculations may eventually be
possible~\cite{Ca1954,Ni1998,SiTe2005,Si2008pra,NeSi2009}, but they rely on
additional mathematical relations fulfilled by the Hamiltonian at hand and
cannot be generalized as easily. 
When the method is
generalized to more complicated configurations~\cite{Si2008pra}, one has to
resort to additional approximations such as the neglect to of terms
proportional to the square of the field strengths [see the text preceding
Eq.~(21) of Ref.~\cite{Si2008pra}]. The latter terms, however, are important in
higher-order Lamb shift calculations~\cite{PaCzJeYe2005,JeCzPa2005}. 

This situation raises a pertinent question.  Let us suppose now that $\{H, J\}
= 0$ and that all the conditions for the application of the chiral method are
fulfilled (Sec.~\ref{sec2C}).  In that case, the outcome of both the standard
as well as the chiral method are unitary transformations.  One could argue that
the ``input'' Hamiltonian, as well as the ``output'' Hamiltonians of the chiral
and standard methods, are Hermitian Hamiltonians connected via a unitary
transformation: They should be equivalent.  Why, then, would the results
contain conflicting terms and fulfill conflicting symmetry relations?  The
reason is to be found in the transformation $U_2$, defined in Eq.~\eqref{U1U2},
which breaks parity and constitutes a chiral transformation which fundamentally
alters the symmetry properties of the Hamiltonian (see also~\ref{appa}).  This
leads to problems, both in regard to the physical interpretation of the
transformed wave functions, and also, in terms of the symmetry properties of
the transformed operators.

While the ``chiral'' transformed Hamiltonian thus is Hermitian and 
is obtained from the ``input'' Hamiltonian 
by a unitary transformation, the operators are 
obtained in nonequivalent representations.
Unfortunately, this implies that the 
realm of applicability of the 
otherwise elegant and concise ``chiral'' Foldy--Wouthuysen
transformation, which is described in Sec.~\ref{sec2C}, 
remains very limited.
For the free Dirac Hamiltonian, the result of the 
chiral method coincides with the one obtained 
using the standard approach (while of course the 
wave function still receives a nontrivial transformation,
see Refs.~\cite{SiTe2005,NeSi2009} and~\ref{appa}). 
For more complicated 
``input'' Hamiltonians, spurious terms are obtained
which break fundamental physical symmetries of the system.
It appears as though the standard approach to the Foldy--Wouthuysen
transformation, while technically more involved and 
perhaps less elegant than the 
``chiral'' or ``exact'' approach, 
remains the most reliable {\em ansatz} for 
the relativistic corrections
which result from a generalized Dirac Hamiltonian.

Beside these considerations, which aim to clarify the 
formal properties of chiral, unitary transformations 
in the context of generalized Dirac Hamiltonians,
we here obtain two results which, to the best of our knowledge,
have not appeared in the literature before.
The first of these concerns the Foldy--Wouthuysen transformation of the 
Dirac Hamiltonian with a scalar potential, given in 
Eq.~\eqref{HSFW_SP}, which contains a somewhat surprising 
anticommutator term proportional to $\{ \vec p^{\,2}, 1/r\}$
which is not present in the Dirac--Coulomb case.
The result given in Eq.~\eqref{HSFW_SP} exhibits
particle-antiparticle symmetry (global prefactor $\beta$).
The same is true for a confining, scalar potential 
[see Eq.~\eqref{HSFW_LC}].
We also give a compact analytic formula
for the leading fourth-order (in the momenta) relativistic corrections to the 
Dirac Hamiltonian in a non-inertial and rotating frame
[see Eq.~\eqref{HSFW_NF}].

%
% Acknowledgments
%
\section*{Acknowledgments}

Support by the National Science Foundation (Grant PHY-1068547)
and by the NIST
(Precision Measurement Grant) is gratefully acknowledged.
The authors acknowledge helpful conversations with 
Istv\'{a}n N\'{a}ndori.

\appendix 

%
% Parity Violation and Spin Operators
%
\section{Parity Violation and Spin Operators}
\label{appa}

The results of the chiral Foldy--Wouthuysen transformations
in Eqs.~\eqref{HCFW_SP},~\eqref{HCFW_DS}, and~\eqref{HCFW_NF}
contain terms which manifestly break parity symmetry.
In order to understand this phenomenon,
we first note that  matrix elements are only invariant under unitary 
transformations if the wave functions also is transformed, according to the 
formula
\begin{align} 
M_{\psi\,\phi} =& \; \left< \psi \left| H \right| \phi \right> = 
\left< \calU \, \psi \left| 
\calU \, H \, \calU^\plus \right| \calU \phi \right> =
\left< \psi' \left|  H' \right| \phi' \right> \,,
\end{align} 
where 
$|\psi'\rangle = \calU \, |\psi\rangle$,
$|\phi'\rangle = \calU \, |\phi\rangle$
and
$H' =  \calU \, H \, \calU^\plus$.
In our case, the transformation $\calU = 
U_2 = \exp(-\ii \, \tfrac{\pi}{4} \, \gamma^5)$
does not correspond to a spinor transformation 
which can be reached from the identity transformation,
within the proper orthochronous Lorentz group:
It is a chiral transformation which changes the 
physical interpretation of the wave function and 
alters the symmetry properties of the Hamiltonian.
Related problem have been considered in the context 
of gauge transformations of 
atomic transition rates, with regard to the 
``length'' and ``velocity'' gauge forms of the 
interaction~\cite{LaRe1950,Ko1978prl,ScBeBeSc1984,EvJeKe2004}
(see also the now famous remark on p.~268 of Ref.~\cite{La1952}).

Let us supplement this argument by some remarks
on the unitary transformations of the operators, 
which also clarify the relation of the transformations 
to the original paper~\cite{FoWu1950}.
The unitary operator which transforms
the free Dirac Hamiltonian into the Foldy--Wouthuysen 
form, can be expressed in closed form as 
follows~\cite{FoWu1950,BjDr1964},
\begin{equation}
U^{\rm (SFW)} = \frac{E_p + m + \beta \, \vec \alpha \cdot \vec p}%
{\sqrt{2 E_p \, (E_p + m)}} \,,
\qquad
E_p = \sqrt{\vec p^{\,2} + m^2} \,.
\end{equation}
One has the relation $U^{\rm (SFW)} \, 
(\vec \alpha \cdot \vec p + \beta m) \,
[U^{\rm (SFW)}]^\plus  = \beta \, E_p$.
The spin matrix transforms as follows,
\begin{align}
U^{\rm (SFW)} \; \vec\Sigma \; [U^{\rm (SFW)}]^\plus =& \;
\vec\Sigma +
\frac{\ii \; \beta}{E_p} \; (\vec \alpha \times \vec p) 
- \frac{1}{E_p \, (E_p + m)} \; 
\left[ \vec p \times \left( \vec\Sigma \times \vec p \right)\right] \,.
\end{align}
This equation corresponds to the entries in 
row~7 of Table~I of Ref.~\cite{FoWu1950}.
We note that $U^{\rm (SFW)} \; \vec\Sigma \; 
[U^{\rm (SFW)}]^\plus$ is a Hermitian operator. 
In particular, we have
$[\ii \; \beta \;  (\vec \alpha \times \vec p)]^\plus = 
-\ii \;  (\vec \alpha \times \vec p)] \; \beta = 
\ii \;  \beta (\vec \alpha \times \vec p)]$.
The inverse operator $[U^{\rm (SFW)}]^\plus$ is obtained 
from $U^{\rm (SFW)}$ by the replacement $\ii \to -\ii$. Hence,
\begin{equation}
U^{\rm (SFW)} \; \left( \vec\Sigma -
\frac{\ii \; \beta}{E_p} \; (\vec \alpha \times \vec p) 
- \frac{1}{E_p \, (E_p + m)} \; 
\left[ \vec p \times \left( \vec\Sigma \times \vec p \right)\right]
\right) \; [U^{\rm (SFW)}]^\plus 
= \vec\Sigma  \,.
\end{equation}
One defines, according to row~11 of Table~I of Ref.~\cite{FoWu1950},
the ``mean'' spin operator as 
\begin{align}
\vec\Sigma_{\rm mean} =& \;
\vec\Sigma -
\frac{\ii \; \beta}{E_p} \; (\vec \alpha \times \vec p) 
- \frac{1}{E_p \, (E_p + m)} \; 
\left[ \vec p \times \left( \vec\Sigma \times \vec p \right)\right] \,.
\end{align}
We then have the relation,
$U^{\rm (SFW)} \; \vec\Sigma_{\rm mean} \; [U^{\rm (SFW)}]^\plus 
= \vec\Sigma$,
i.e.~the spin matrix $\vec \Sigma$ can be identified 
as the relevant physical spin operator in the Foldy--Wouthuysen
representation.
The identification as a ``mean'' operator is motivated by the 
fact that, in the nonrelativistic picture, 
the zitterbewegung term describes the influence of the 
``quiver'' motion of the electron on its quantum 
trajectory in the mean, i.e., without the 
influence of the instantaneous velocity operator
which has the eigenvalues of the $\vec\alpha$ matrix.
The leading nonrelativistic kinetic term 
is $\vec p^{\,2}/(2 m)$, which implies that
all the operators after the Foldy--Wouthuysen transformation
can be interpreted as ``mean'' operators.
We note that the all terms in the mean spin operator
$\vec\Sigma_{\rm mean}$ constitute pseudovectors.
Under parity, we have
$\vec \Sigma \to \vec \Sigma$, $\vec p \to -\vec p$,
and $\vec \alpha \to -\vec \alpha$.
The total angular momentum
$\vec J = \vec L + \tfrac12 \, \vec\Sigma$
commutes with the Dirac Hamiltonians investigated 
here in Eqs.~\eqref{HDC},~\eqref{HSP},~\eqref{HDS} 
and~\eqref{HNF}, and with their (standard)
Foldy--Wouthuysen transforms given in 
Eqs.~\eqref{HSFW_DC},~\eqref{HSFW_SP},~\eqref{HSFW_LC},~\eqref{HSFW_DS}
and~\eqref{HSFW_NF}.

One may transform the spin operator 
$\vec \Sigma$ into the Eriksen--Kolsrud (``chiral'')
representation~\cite{ErKo1958}, using the unitary transformation 
$U^{\rm (CFW)} \, [U^{\rm (SFW)}]^{-1}$.
In this case, one obtains, according to 
Eq.~(10) of Ref.~\cite{SiTe2005}, 
terms proportional to $\vec p \times \vec \Sigma$, 
which transform as vectors under parity
($\vec p \to - \vec p$, but $\vec \Sigma \to \vec \Sigma$).
This implies that the spin operator in the 
chiral representation cannot be consistently identified 
with the spin matrix $\vec\Sigma$, not even in 
the nonrelativistic limit. This consideration affords an 
alternative view on the generation of the parity-breaking terms.

\end{document}